\documentclass[a4paper]{jpconf}
\usepackage{graphicx}
\newcommand{\ba}{\begin{eqnarray}}
\newcommand{\ea}{\end{eqnarray}}

\begin{document}

\title{Electromagnetic couplings of pentaquarks}

\author{E. Ortiz-Pacheco and R. Bijker}
\address{Instituto de Ciencias Nucleares, 
Universidad Nacional Aut\'onoma de M\'exico, 
A.P. 70-543, 04510 Ciudad de M\'exico, M\'exico}
\ead{bijker@nucleares.unam.mx}

\begin{abstract}
In this contribution, we discuss the electromagnetic couplings of pentaquark states with hidden charm. 
This work is motivated by recent experiments at CERN by the LHCb Collaboraton and current experiments 
at JLab to confirm the existence of hidden-charm pentaquarks in photoproduction experiments.
\end{abstract}
 
\section{Introduction}

The observation of hidden-charm pentaquark states by the LHCb Collaboration \cite{LHCb1}-\cite{LHCb4} 
has created an avalanche of theoretical studies on the nature of these states and on different plausible 
interpretations of the observed signals, {\it e.g.} kinematical effects \cite{nonresonant1,nonresonant2,nonresonant3}, 
molecular states \cite{Karliner}-\cite{Yamaguchi} and compact pentaquarks \cite{Maiani}-\cite{JPG2019}. 
The existence of narrow $N^*$ and $\Lambda^*$ resonances with hidden charm was predicted \cite{Oset1,Oset2} 
in a coupled-channel unitary approach five years before the LHCb data were published, as well as in Refs.~\cite{YangZC}. 
More information on the experimental and theoretical aspects of pentaquark states,  
as well as a more complete list of references, can be found in the reviews \cite{review1}-\cite{review6}. 

In the present contribution we discuss the electromagnetic couplings of $uudc\bar{c}$ hidden-charm 
pentaquark states which are relevant for photoproduction experiments at JLab 
\cite{photo1}-\cite{photo6}.

\section{Pentaquark states}

Pentaquark states depend both on the orbital degrees of freedom and the internal degrees of freedom of color, 
spin and flavor
\ba
\psi \;=\; \psi^{\rm o} \psi^{\rm c} \phi^{\rm f} \chi^{\rm s} ~.
\ea
The construction of the classification scheme of $uudc\bar{c}$ pentaquark states was carried out expicitly in Ref.~\cite{JPG2019,Cocoyoc2017} using the following two conditions: (i) the pentaquark wave function should be 
a color singlet and (ii) the wave function of the four-quark subsystem should be antisymmetric. 
The permutation symmetry of four-quark states is characterized by the $S_4$ Young tableaux $[4]$, $[31]$, 
$[22]$, $[211]$ and $[1111]$ or, equivalently, by the irreducible representations of the tetrahedral group 
${\cal T}_d$ (which is isomorphic to $S_4$) as $A_1$, $F_2$, $E$, $F_1$ and $A_2$, respectively. 

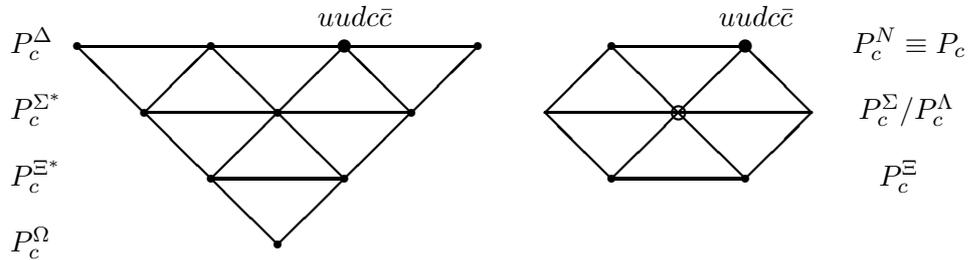
\begin{figure}
\centering
\setlength{\unitlength}{0.5pt}
\begin{picture}(660,250)(0,0)
\thicklines
\put(230,215){$uudc\bar{c}$}
\put( 50,200) {\line(1,0){300}}
\put(100,150) {\line(1,0){200}}
\put(150,100) {\line(1,0){100}}
\put(200, 50) {\line(1,1){150}}
\put(150,100) {\line(1,1){100}}
\put(100,150) {\line(1,1){ 50}}
\put(200, 50) {\line(-1,1){150}}
\put(250,100) {\line(-1,1){100}}
\put(300,150) {\line(-1,1){ 50}}
\put(250,200){\circle*{10}}
\multiput( 50,200)(100,0){4}{\circle*{5}}
\multiput(100,150)(100,0){3}{\circle*{5}}
\multiput(150,100)(100,0){2}{\circle*{5}}
\put(200, 50){\circle*{5}}
\put(  0,195) {$P_c^{\Delta}$}
\put(  0,145) {$P_c^{\Sigma^{\ast}}$}
\put(  0, 95) {$P_c^{\Xi^{\ast}}$}
\put(  0, 45) {$P_c^{\Omega}$}
\put(530,215){$uudc\bar{c}$}
\put(450,200) {\line(1,0){100}}
\put(400,150) {\line(1,0){200}}
\put(450,100) {\line(1,0){100}}
\put(400,150) {\line(1,1){ 50}}
\put(450,100) {\line(1,1){100}}
\put(550,100) {\line(1,1){ 50}}
\put(450,100) {\line(-1,1){ 50}}
\put(550,100) {\line(-1,1){100}}
\put(600,150) {\line(-1,1){ 50}}
\put(550,200){\circle*{10}}
\multiput(450,200)(100,0){2}{\circle*{5}}
\multiput(4000,150)(100,0){3}{\circle*{5}}
\multiput(450,100)(100,0){2}{\circle*{5}}
\put(500,150){\circle{10}}
\put(630,195) {$P_c^{N} \equiv P_c$}
\put(635,145) {$P_c^{\Sigma}/P_c^{\Lambda}$}
\put(650, 95) {$P_c^{\Xi}$}
\end{picture}
\caption{Pentaquark decuplet and octet}
\label{flavor}
\end{figure}

The first condition that the pentaquark wave function has to be a color-singlet, implies that the color wave 
function of the four-quark configuration has to be a $[211]$ triplet with $F_1$ symmetry under ${\cal T}_d$.  
As a consequence, the second condition that the total $q^4$ wave function has to be antisymmetric ($A_2$), 
means that the orbital-spin-flavor part is a $[31]$ triplet with $F_2$ symmetry 
\ba
\psi \;=\; \left[ \psi^{\rm c}_{F_1} \times \psi^{\rm osf}_{F_2} \right]_{A_2} ~,
\label{pentaquarkwf}
\ea
where the subindices refer to the symmetry properties of the four-quark subsystem under permutation. Moreover, 
in this contribution we limit ourselves to ground-state pentaquark states, {\it i.e.} without orbital excitations, 
which are symmetric ($A_1$). Therefore, the spin-flavor part is a $[31]$ state with $F_2$ symmetry  
\ba
\psi \;=\; \psi^{\rm o}_{A_1} \left[ \psi^{\rm c}_{F_{1}} \times \psi^{\rm sf}_{F_{2}} \right]_{A_2} ~. 
\ea
In Ref.~\cite{JPG2019} it was shown that there are in total seven $uudc\bar{c}$ ground-state pentaquark 
configurations with angular momentum and parity $J^P=3/2^-$ (which is quoted in the literature as the most likely 
value of the angular momentum and parity of the $P_c$ pentaquark \cite{LHCb1}), three of which belong to a flavor decuplet 
and the remaining four to a flavor octet (see Fig.~\ref{flavor} and first column of Table~\ref{photo}).

\section{Electromagnetic couplings}

For experiments that aim to study pentaquarks through near threshold $J/\psi$ photoproduction 
at JLab, the size of the electromagnetic couplings of the pentaquarks is important. Here we 
discuss the electromagnetic couplings for the ground state pentaquarks with spin and parity 
$J^P=3/2^-$. Electromagnetic couplings are described by 
\ba
{\cal H}_{\rm em} \;=\; e \int d^3 x \, \hat{J}^{\mu}(\vec{x}) A_{\mu}(\vec{x}) ~,
\label{hem}
\ea
where $J^{\mu}$ is the electromagnetic current summed over all quark flavors 
\ba
\hat{J}^{\mu}(\vec{x}) \;=\; \sum_q e_q \, \bar{q}(\vec{x}) \gamma^{\mu} q(\vec{x}) ~.
\ea

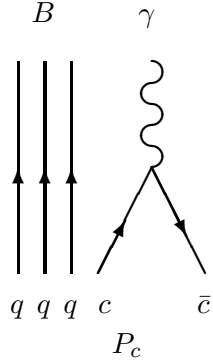
\begin{figure}
\centering
\setlength{\unitlength}{1pt}
\begin{picture}(100,150)(0,0)
\thicklines
\put( 70, 84) {\oval(8,8)[r]}
\put( 70, 92) {\oval(8,8)[l]}
\put( 70,100) {\oval(8,8)[r]}
\put( 70,108) {\oval(8,8)[l]}
\put( 70,116) {\oval(8,8)[r]}
\put( 20,40) {\line(0,1){80}}
\put( 30,40) {\line(0,1){80}}
\put( 40,40) {\line(0,1){80}}
\put( 50,40) {\line( 1,2){20}}
\put( 90,40) {\line(-1,2){20}}
\put( 20,40) {\vector(0,1){40}}
\put( 30,40) {\vector(0,1){40}}
\put( 40,40) {\vector(0,1){40}}
\put( 50,40) {\vector(1, 2){10}}
\put( 70,80) {\vector(1,-2){13}}
\put( 17,25) {$q$}
\put( 27,25) {$q$}
\put( 37,25) {$q$}
\put( 50,25) {$c$}
\put( 87,25) {$\bar{c}$}
\put( 55,10) {$P_c$}
\put( 25,135) {$B$}
\put( 65,135) {$\gamma$}
\end{picture}
\caption{Electromagnetic decay of pentaquark 
$P_c$ into a baryon $B$ and a photon, 
$P_c \rightarrow B + \gamma$.}
\label{emdecay}
\end{figure}

The electromagnetic coupling of Eq.~(\ref{hem}) describes both the emission (and absorption) of 
the photon off a quark, as is used in studies of photocouplings of baryons \cite{Copley}, and 
the annihilation process of a quark-antiquark pair~\cite{LeYaouanc}. For the process of interest, 
$P_c \rightarrow N + \gamma$, the relevant term is the annilation of a pair of $c\bar{c}$ quarks, 
Fig.~\ref{emdecay}. In the present calculation we use the nonrelativistic form of the interaction. 
The radiative decay widths can be calculated as \cite{BIL2}
\ba
\Gamma(P_c \rightarrow N + \gamma) \;=\; 
\frac{\rho }{(2\pi)^2} \, \frac{2}{2J+1} \sum_{\nu>0} | A_{\nu}(k) |^2 ~, 
\label{gw}
\ea
where $\rho$ is the phase space factor, and $A_{\nu}$ denotes the helicity amplitude 
\ba
A_{\nu}(k) \;=\; \left< N,1/2^+,\nu-1;\gamma \left| {\cal H}_{\rm em}^{\rm nr} \right| P_c,3/2^-,\nu \right>
\;=\; \sqrt{\frac{4\pi\alpha}{k_0}} \, \beta_{\nu} \, F(k) ~.
\ea
Here $\alpha$ is the fine-structure constant, $k_0$ and $k=|\vec{k}|$ represent the energy and the momentum of the photon. The coefficient $\beta_{\nu}$ is the contribution from the color-spin-flavor part for the annihilation of a $c\bar{c}$ color-singlet pair with spin $S=S_z=1$. It is straightforward to show that for the cases considered, {\it i.e.} ground-state pentaquarks with $J^P=3/2^-$, the helicity amplitudes are related by
\ba
\beta_{1/2} \;=\; \beta_{3/2}/\sqrt{3} ~.
\ea

\begin{table}[b]
\centering
\caption{Contribution from the color-spin-flavor part to the helicity amplitudes for the electromagnetic 
decays of $uudc\bar{c}$ decuplet (top) and octet (bottom) pentaquark states into $N + \gamma$. Here $e_c$ 
is the electric charge of the charm quark $e_c=2/3$.}
\label{photo}
\vspace{10pt}
\begin{tabular}{cccc}
\hline
\noalign{\smallskip}
State & Name & $\beta_{1/2}$ & $\beta_{3/2}$ \\
\noalign{\smallskip}
\hline
\noalign{\smallskip}
$\left[ \phi_{A_1} \times \chi_{F_2} \right]_{F_2}$ & $P_c^{\Delta}$ 
& $0$ & $0$ \\
\noalign{\smallskip}
$\left[ \phi_{F_2} \times \chi_{A_1} \right]_{F_2}$ & $P_c^{\Delta}$ 
& $0$ & $0$ \\
\noalign{\smallskip}
$\left[ \phi_{F_2} \times \chi_{F_2} \right]_{F_2}$ & $P_c^{\Delta}$ 
& $0$ & $0$ \\
\noalign{\smallskip}
\hline
\noalign{\smallskip}
$\left[ \phi_{F_2} \times \chi_{A_1} \right]_{F_2}$ & $P_c^{N}$ 
& $0$ & $0$ \\
\noalign{\smallskip}
$\left[ \phi_{F_2} \times \chi_{F_2} \right]_{F_2}$ & $P_c^{N}$ 
& $\phantom{-}\frac{1}{6\sqrt{2}} e_c$ & $\phantom{-}\frac{1}{2\sqrt{6}} e_c$ \\
\noalign{\smallskip}
$\left[ \phi_{E} \times \chi_{F_2} \right]_{F_2}$ & $P_c^{N}$ 
& $-\frac{1}{6} e_c$ & $-\frac{1}{2\sqrt{3}} e_c$ \\
\noalign{\smallskip}
$\left[ \phi_{F_1} \times \chi_{F_2} \right]_{F_2}$ & $P_c^{N}$ 
& $-\frac{1}{2\sqrt{6}} e_c$ & $-\frac{1}{2\sqrt{2}} e_c$ \\
\noalign{\smallskip}
\hline
\end{tabular}
\end{table}

Finally, $F(k)$ is a form factor denoting the contribution from the orbital part of the pentaquark wave function. Its specific form depends on the type of quark model used: harmonic oscillator, hypercentral, or other. Here we concentrate on the color-spin-flavor part which is common to all quark models. In Table~\ref{photo} we show the results for the contribution from the color-spin-flavor part to the helicity amplitudes for different configurations of $uudc\bar{c}$ pentaquarks. The couplings to the octet configuration with $\phi_{A_1}$ and the three decuplet configurations vanish because of symmetry reasons. The strongest coupling is to the octet pentaquark configuration with $\phi_{F_1}$, followed by $\phi_{E}$ and $\phi_{F_2}$.  

\section{Summary and conclusions}

In conclusion, in this contribution we discussed the electromagnetic couplings of ground-state $uudc\bar{c}$ 
pentaquark states with angular momentum and parity $J^P=3/2^-$. At present we did not include orbital excitations. 
Of the seven possible configurations only three octet configurations have a nonvanishing photocoupling.  
Since the photon momentum is large, we expect that these couplings are strongly suppressed by the form factor, 
$F(k)$, which represents the contribution from the orbital part of the pentaquark wave function.  

If the signal observed by the LHCb Collaboration indeed corresponds to hidden-charm pentaquarks, there should 
be an entire multiplet of pentaquark states, for example a pentaquark octet consisting of $P_c^N$, $P_c^{\Sigma}$, 
$P_c^{\Lambda}$ and $P_c^{\Xi}$ states. 

\ack
This work was supported in part by grant IN109017 from DGAPA-UNAM, Mexico
and grants 251817 and 340629 from CONACyT, Mexico.

\end{document}